\documentclass[12pt]{iopart}
\usepackage{eufrak}
\usepackage{graphicx}

\begin{document}
\title{Squeezing of open boundaries by Maxwell-consistent real
coordinate transformation}
\author{D~M~Shyroki}
\address{Department of Communications, Optics and Materials,
Technical University of Denmark, Building 343v, 2800 Kgs.~Lyngby,
Denmark} \ead{ds@com.dtu.dk}

\begin{abstract}
To simulate open boundaries within finite computation domain,
real-function coordinate transformation in the framework of
generally covariant formulation of Maxwell equations is proposed.
The mapping---realized with arctangent function here---has a
transparent geometric meaning of pure squeezing of space, is
admissible by classical electrodynamics, does not introduce
artificially lossy layers (or `lossy coordinates') to absorb
outgoing radiation nor leads to non-Maxwellian fields. At the same
time, like for anisotropic perfectly matched layers, no
modification (except for transformation of material tensors) is
needed to existing nearest-neighbor computation schemes, which
makes it well suited for parallel computing implementation.
\end{abstract}

\section{Introduction}
Direct numerical methods of modeling electromagnetic phenomena,
such as finite-difference time-domain (FDTD) and frequency-domain
(FDFD) schemes, are invariably concerned with how to represent
infinite space surrounding the region of interest on a bounded
computation domain. Two approaches to combat that problem do
exist; one, as in~\cite{Lindman75} or~\cite{Higdon87}, can be
classified as non-local since it is not limited to the treatment
of nearest-neighbor interactions on a computation grid due to
higher-order differentials appearing in the formulation, a
shortcoming when parallel computer implementation is considered;
another is aimed to modify the (local) material properties of
boundary regions in such a way that outgoing radiation experiences
no parasitic reflections from the boundaries of computation
window~\cite{Rappaport92, Sacks95}, and hence can be called local.

The now-classic local technique to represent open boundaries
(called absorbing boundaries when zero reflectivity of surrounding
space is emphasized and mimicked in simulations) is by means of
absorbing perfectly matched layers (PMLs) introduced originally
in~\cite{Berenger94}. The technique, especially in its non-split
version~\cite{Sacks95}, is considered simple and efficient, and
enjoy great popularity in the electromagnetic modeling community.
Nonetheless it cannot be considered as completely perfect because:
\begin{enumerate}
    \item free PML parameters, such as maximum
    conductivity and conductivity profile, bear not always obvious
    geometric or physical relation to particular problem to
    simulate, hence need for the adjustment and optimization of
    PMLs so often;
    \item use of complex-valued matrices for modified dielectric
    permittivity $\epsilon$ and magnetic permeability $\mu$ leaves
    no room for CPU time and memory savings
    with real-field finite-difference formulations;
    \item loss of accuracy when solving eigenproblems in
    frequency domain is inevitable (though normally minor)
    due to nonzero mode tails extending into the regions of
    modified $\epsilon$ and $\mu$ within PMLs;
    \item difficulties with non-Cartesian and, in particular,
    non-orthogonal grids have been reported~\cite{Buksas01}.
\end{enumerate}
As an attempt to overcome these problems while retaining locality
of the formulation, we propose a conceptually simple and
numerically easy-to-implement squeezing of open boundaries (SOB)
technique in this Letter.

The underlying idea of SOB is to map infinite surrounding space
(or rather the whole space, with better sampling for central
region and coarser for outskirts) onto the finite computation
domain, instead of inserting anisotropic absorbing PMLs between
the region of interest and computation boundaries. A clever way to
do such mapping inexpensively is by transforming $\epsilon$ and
$\mu$ fields as stipulated by generally covariant electrodynamics,
while retaining the form of Maxwell equations untouched. The
mapping---illustrated by use of arctangent function here, with
other possibilities discussed---is rigorous at the stage of
analytic description; is smooth, while the $n$th derivative of
material tensors at PML interface (with $n$ depending on the order
of the profile) is discontinuous; and is real-valued, enabling
finite-difference algorithms in real notation where appropriate.
Another advantage of the SOB method is its extendability: that is,
anisotropic, magnetic materials and nontrivial backgrounds can be
treated straightforwardly; and with same ease non-Cartesian and
non-orthogonal coordinates, if preferred, can be squeezed in the
manner proposed. Finally, this technique justifies a surprising
possibility for lossless PML formulation.

\section{Covariant Maxwell equations}
It was Lorentz-covariance of Maxwell equations that led to special
relativity over a century ago. Another, less celebrated though
well and long ago established fact about Maxwell equations is that
they can be formulated in a generally covariant manner, i.e., so
that they do not change their form under arbitrary reversible
transformation from Cartesian coordinates~\cite{SchoutenBOOK1951,
PostBOOK1962}. Surprisingly, this feature was first exploited in
direct computation electromagnetics perhaps only a decade ago, in
`logically Cartesian' FDTD simulations of high index contrast
dielectric structures~\cite{WardPendry96, WardPendry98} (see
also~\cite{ShyrokiOnWardPendry}). For the sake of completeness let
us write coordinate-invariant Maxwell equations here, in terms of
electric covariant vector $E_\nu$ and magnetic covariant
pseudovector $\tilde{H}_\nu$ as in~\cite{ShyrokiOptLett06}:
\begin{eqnarray}
    \tilde\mathfrak{E}^{\kappa\lambda\nu}\partial_\lambda E_\nu =
    -\mu^{\kappa\lambda}\dot{\tilde{H}}_\lambda, &\quad&
    \partial_{\kappa}\mu^{\kappa\lambda}\tilde{H}_\lambda = 0,
    \label{MaxwellInvMod1}\\
    \tilde\mathfrak{E}^{\kappa\lambda\nu}\partial_\lambda \tilde{H}_\nu
    = \epsilon^{\kappa\lambda} \dot{E}_\lambda + \mathfrak{j}^\kappa,
    &\quad& \partial_{\kappa}\epsilon^{\kappa\lambda} E_\lambda = \rho,
    \label{MaxwellInvMod2}
\end{eqnarray}
a form which, written in components explicitly, is identical to
conventional Cartesian representation (pseudo permutation field
$\tilde\mathfrak{E}^{\kappa\lambda\nu}$ equals Levi-Civita symbol
in any coordinate system), with the constitutive relations
\begin{eqnarray}\label{MaterialEqsInv}
    \tilde\mathfrak{B}^\lambda =
    \mu^{\lambda\nu}\tilde{H}_\nu, \quad
    \mathfrak{D}^\lambda = \epsilon^{\lambda\nu} E_\nu
\end{eqnarray}
(i.e., no optical activity assumed), where
$\tilde\mathfrak{B}^\lambda$ is magnetic induction pseudo vector
density of weight $+1$, and $\mathfrak{D}^\lambda$ is electric
induction vector density; hence $\epsilon^{\lambda\nu}$ and
$\mu^{\lambda\nu}$ are contravariant tensor densities transformed
according to
\begin{equation}\label{MaterialConstantsTransform}
    \epsilon^{\lambda\nu} = |\Delta|^{-1} J_{\lambda'}^{\lambda}
    J_{\nu'}^{\nu} \epsilon^{\lambda'\nu'}, \quad
    \mu^{\lambda\nu} = |\Delta|^{-1} J_{\lambda'}^{\lambda}
    J_{\nu'}^{\nu} \mu^{\lambda'\nu'},
\end{equation}
where $J_{\lambda'}^{\lambda} \equiv
\partial_{\lambda'}x^{\lambda}$ is the Jacobian transformation
matrix for contravariant components, $\Delta \equiv \det
J_{\lambda'}^{\lambda}$ is its determinant.

Such formulation enables one to hide all metric information into
$\epsilon^{\lambda\nu}$ and $\mu^{\lambda\nu}$ while invariably
using Cartesian-like representation of Maxwell equations
(\ref{MaxwellInvMod1}), (\ref{MaxwellInvMod2}), which is extremely
convenient; no wonder many authors strived (successfully) to
`derive' (\ref{MaterialConstantsTransform}) on different grounds
and under different assumptions, as in~\cite{WardPendry96} for
isotropic media or in~\cite{Teixeira98} for diagonal
transformation matrices $J_{\lambda'}^{\lambda}$. It is worth
noting that, however, (\ref{MaterialConstantsTransform}) are a
direct consequence of transformation characteristics assigned to
electric and magnetic fields; hence their general nature and no
need in any tricky derivations. In practice, dielectric
permittivity and magnetic permeability are referenced to Cartesian
frame, so if one wants to use non-Cartesian coordinates instead,
with Cartesian-like equations (\ref{MaxwellInvMod1}),
(\ref{MaxwellInvMod2}), then transformation rules
(\ref{MaterialConstantsTransform}) are to be employed and
specified for transformation from Cartesian to those coordinates.
We make such specification for the mapping onto
arctangent-squeezed coordinates in the next Section.

\section{Open boundaries on confined domain}\label{sec_arctanFormulation}
Let the Cartesian coordinates $\{x^{1'},x^{2'},x^{3'}\} =
\{x,y,z\}$ be transformed to $\{x^{1},x^{2},x^{3}\} = \{u,v,w\}$
according to
\begin{equation}\label{arctanTransform}
    u = \arctan(x/x_0), \quad
    v = \arctan(y/y_0), \quad
    w = \arctan(z/z_0).
\end{equation}
where $x_0$, $y_0$, $z_0$ are the units of length along the
corresponding coordinates. Such mapping preserves central region
of space (where the scatterer or the waveguide is supposedly
located) virtually untouched, while smoothly squeezing the outer
space into the $(-\pi/2,\pi/2)^3$ bounded computation domain. The
$x_0$, $y_0$ and $z_0$ units are arbitrary at this, analytic
stage, but their careless choice may compromise accuracy of
finite-difference calculations on (more or less) equidistant grids
in squeezed coordinates. Indeed, poor sampling on a `squeezed'
grid of `physical' lengths far from the origin of coordinates
($|x|\gg x_0$, $|y|\gg y_0$, or $|z|\gg z_0$) is a reason to match
scaling factors $x_0$, $y_0$ and $z_0$ with actual physical
dimensions of the region of interest or with the wavelength. Note
that in the proposed approach, $x_0$, $y_0$ and $z_0$ are the only
parameters to be adjusted to particular physical problem, with a
clear geometric relation to that problem.

By differentiating (\ref{arctanTransform}) one gets the
transformation matrix
\begin{equation}\label{Amatrix}
    J_{\lambda'}^{\lambda} = \left(%
\begin{array}{ccc}
  x_0^{-1}\cos^2 u & 0 & 0 \\
  0 & y_0^{-1}\cos^2 v & 0 \\
  0 & 0 & z_0^{-1}\cos^2 w \\
\end{array}%
\right).
\end{equation}
The `squeezed' permittivity and permeability can be obtained with
(\ref{MaterialConstantsTransform}), (\ref{Amatrix}) immediately;
the diagonal components of $\epsilon^{\lambda\nu}$, for example,
are
\begin{equation}\label{EpsTransform}
\fl    \epsilon^{uu} = \epsilon^{xx} \frac{y_0 z_0 \cos^2 u}{x_0
    \cos^2 v\cos^2 w}, \quad
    \epsilon^{vv} = \epsilon^{yy} \frac{z_0 x_0 \cos^2 v}{y_0
    \cos^2 u\cos^2 w}, \quad
    \epsilon^{ww} = \epsilon^{zz} \frac{x_0 y_0 \cos^2 w}{z_0
    \cos^2 u\cos^2 v},
\end{equation}
and similarly for $\mu^{\lambda\lambda}$. It should be emphasized
that even if matrix representations of dielectric permittivity and
magnetic permeability are non-diagonal in Cartesian coordinates,
this generates no additional complexity in deriving the
$\epsilon^{\lambda\nu}$ and $\mu^{\lambda\nu}$ off-diagonal
components; here we omit them for brevity. Specifying the
transformed $\epsilon^{\lambda\nu}$ and $\mu^{\lambda\nu}$ for
squeezed cylindrical or spherical coordinates also poses no
difficulty in present approach, but is out of scope here.

The structure of (\ref{EpsTransform}) resembles that of modified
$\epsilon$ and $\mu$ in PML regions with `stretching
variables'~\cite{Chew94} $s_x = J^u_x$ etc., but transformation
(\ref{EpsTransform}) contains no complex-valued functions and is
smooth over the entire spatial domain. The price for the
smoothness is that under arctangent
transformation~(\ref{arctanTransform}), which is linear only in
the vicinity of the origin of coordinates, familiar geometric
figures become distorted on new mesh and, analytically, are
defined by equations with $x$, $y$ and $z$ coordinates replaced by
$x_0\tan u$ etc; this poses high demands on index averaging
technique at index discontinuities for better numerical
convergence, as found in Section~\ref{FDFD_simulations} for
frequency-domain eigenproblem. How to go around that by modifying
the transformation function in~(\ref{arctanTransform}) is
explained in Section~\ref{alternative_transforms}, where we end up
with non-absorbing PML formulation.

\section{FDFD simulation of guided modes}\label{FDFD_simulations}
\begin{figure}
\centering
\includegraphics[width=2.5in]{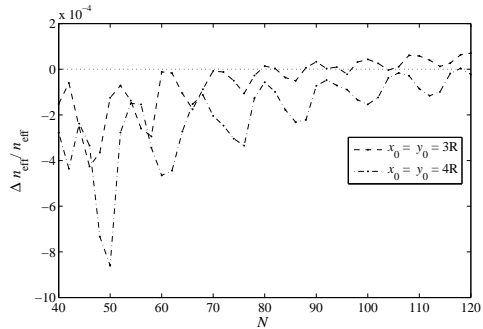}
\caption{Relative discrepancy $\Delta n_\mathrm{eff} /
n_\mathrm{eff}$ between SOB-FDFD computed and analytically
calculated mode indices of step-index fiber (radius $R$ = 2
$\mu$m, material index 1.45) in air background at 1.45 $\mu$m,
with increasing the number of grid points $N$ per domain width.}
\label{fig_convergence}
\end{figure}

It is common intuition that light tends to concentrate in
high-index regions, so one may expect that dielectric profile
transformed according to~(\ref{arctanTransform}) shall support
spurious modes guided along computation boundaries where some of
the permittivity and permeability components head to infinity. To
disprove this, guided propagation in step-index fiber was
simulated with full-vector FDFD algorithm implementing
permittivity and permeability profiles as in~(\ref{EpsTransform}),
adopted to the two-dimensional geometry. No spurious
boundary-guided solutions were found, and numeric results show
reasonable agreement with those obtained analytically for the same
fiber (figure~\ref{fig_convergence}). Convergence was found
affected by the scaling parameters $x_0$, $y_0$ (the two curves in
the figure correspond to the $x_0 = y_0 = 3R$ and $x_0 = y_0 = 4R$
choices, where $R$ is the fiber radius), and very sensitive to
index averaging scheme. This latter sensitivity is due to highly
steep and nonlinear profiles of $\epsilon$ and $\mu$ in squeezed
coordinates, which can lead to systematic under- or overestimation
of the eigenvalues if inappropriate index averaging at material
interfaces (e.g., simple volume-weighted averaging, quite
widespread in FDTD and FDFD modeling) is used.
%

\section{A new way to construct PMLs}\label{alternative_transforms}
An interesting question is how arbitrary the transform function
in~(\ref{arctanTransform}) is. Indeed, one may use, e.g.,
hyperbolic tangent for the mapping; such choice would lead to only
slight changes in (\ref{Amatrix}), (\ref{EpsTransform}), with
$\cos^2(u\pi/2)$ etc.~on $(-1, 1)$ substituted by $1 - u^2$ on the
same domain. And indeed, numeric simulations in
$\tanh$-transformed coordinates show results similar to those in
Section~\ref{FDFD_simulations}, although convergence was found
poorer for the example in figure~\ref{fig_convergence}, probably
owing to higher steepness of appropriately transformed $\epsilon$
and $\mu$ profiles near the domain boundaries which spur numerical
errors at discretization.

Going a step further, one might introduce piecewise mapping
functions like $u = x$ for $|x| \leq x_1$, $u = x_1 +
x_2\tanh\frac{x - x_1}{x_2}$ for $|x| > x_1$, where $x_1$ defines
the interface between space region untouched by the
transformation, and $x_2$ the width of `non-absorbing perfectly
matched layer'. The analogy with conventional dispersive PMLs
becomes even more pronounced if we put $x_2 \propto \lambda$,
which is a rather natural choice as noted in
Section~\ref{sec_arctanFormulation}. The advantage of this
`piecewise' formulation over smooth arctangent or hyperbolic
tangent squeezing is that within the $|x| \leq x_1$, $|y| \leq
y_1$, $|z| \leq z_1$ region, $\epsilon$ and $\mu$ profiles are
defined as on untransformed grid; the disadvantage is that they
are consequently sharper near computation domain boundaries, for
the domain of the same width.

The notions of `complex coordinate stretching'~\cite{Chew94} or
`lossy mapping of space'~\cite{Rappaport95} have long been used to
derive standard (lossy) PMLs, though it was not always clear what
physical sense those complex-valued `degrees of freedom' added by
hand to spatial variables in Maxwell equations do have; in
operational terms, whether complex-valued coordinates are
observable. The proposed SOB technique paves the way to construct
lossless PMLs with a clear geometric meaning of outer space
squeezing; and if losses should be introduced (as in
frequency-domain calculations of leaky modes), this can be done by
adding, under the leakage irreversibility condition, an imaginary
part to refractive index of surrounding medium before squeezing.

\section{Conclusion}
Squeezing of open boundaries is proposed as an inexpensive and, at
analytic stage, rigorous alternative to standard lossy PML
technique. What makes our method so attractive is its conceptual
clarity: we do not surround computation window with artificial
lossy media; we do not modify Maxwell equations in any way; all we
do is we choose coordinate system allowed by covariant nature of
Maxwell equations and suitable for calculations---and for the
finite-difference or finite-element calculations on a bounded
domain, a suitable system is one that has bounded coordinates. The
method is more straightforward to apply in time domain; in our
proof-of-principle frequency-domain simulations of guided
propagation, no spurious modes confined in the regions of strongly
modified $\epsilon$ and $\mu$ have been detected.



\section*{References}
\begin{thebibliography}{99}

\bibitem{Lindman75}E.~Lindman, ``Free-space boundary conditions
for the time dependent wave equation,'' \textit{J. Comput. Phys.},
vol. 18, pp. 66--78, 1975.

\bibitem{Higdon87}R.~L.~Higdon, ``Numerical absorbing boundary
conditions for the wave equation,'' \textit{Math. Comput.}, vol.
49, pp. 65--90, 1987.

\bibitem{Rappaport92}C.~M.~Rappaport and L.~Bahrmasel, ``An
absorbing boundary condition based on anechoic absorber for EM
scattering computation,'' \textit{J. Electromag. Waves Appl.},
vol. 6, pp. 1621--1634, 1992.

\bibitem{Sacks95}Z.~S.~Sacks, D.~M.~Kingsland, R. Lee, and J.-F.
Lee, ``A perfectly matched anisotropic absorber for use as an
absorbing boundary condition,'' \textit{IEEE Trans. Antennas
Propagat.}, vol. 43, pp. 1460--1463, 1995.

\bibitem{Berenger94}J.~P.~B\'{e}renger, ``A perfectly matched
layer for the absorption of electromagnetic waves,'' \textit{J.
Comput. Phys.}, vol. 114, pp. 185--200, 1994.



\bibitem{Buksas01}M.~W. Buksas, ``Implementing the perfectly
matched layer absorbing boundary condition with mimetic
differencing schemes,'' \textit{Prog. Electromagn. Research PIER},
vol. 32, pp. 383-–411, 2001.

\bibitem{SchoutenBOOK1951}J.~A. Schouten, \textit{Tensor Analysis
for Physicists} (Clarendon, Oxford, 1951).

\bibitem{PostBOOK1962}E.~J. Post, \textit{Formal Structure of
Electromagnetics} (North-Holland, Amsterdam, 1962).

\bibitem{WardPendry96}A.~J. Ward and J.~B. Pendry, ``Refraction
and geometry in Maxwell's equations,'' \textit{J. Modern Opt.},
vol. 43, pp. 773--793, 1996.

\bibitem{WardPendry98}A.~J. Ward and J.~B. Pendry, ``Calculating
photonic Green's functions using a nonorthogonal finite-difference
time-domain method,'' \textit{Phys. Rev.} B, vol. 58, pp.
7252--7259, 1998.

\bibitem{ShyrokiOnWardPendry}D.~M. Shyroki, ``Note on
transformation to general curvilinear coordinates for Maxwell's
curl equations,'' arXive:physics/0307029, 2003.

\bibitem{ShyrokiOptLett06}D.~M. Shyroki, ``Exact equivalent-profile
formulation for bent optical waveguides,''
arXive:physics/0605002, 2006.

\bibitem{Teixeira98}F.~L. Teixeira and W.~C. Chew, ``General
closed-form PML constitutive tensors to match arbitrary
bianisotropic and dispersive linear media,'' \textit{IEEE
Microwave Guided Wave Lett.}, vol. 8, pp. 223--225, 1998.

\bibitem{Chew94}W.~C. Chew and W.~H. Weedon, ``A 3D perfectly
matched medium from modified Maxwell's equations with stretched
coordinates,'' \textit{Microwave Opt. Tech. Lett.}, vol. 7, pp.
599--604, 1994.

\bibitem{Rappaport95}C.~M. Rappaport, ``Perfectly mathed
absorbing boundary conditions based on anisotropic lossy mapping
of space,'' \textit{IEEE Microwave Guided Wave Lett.}, vol. 5, pp.
90--92, 1995.

\end{thebibliography}
\end{document}